\documentclass[12pt]{iopart}

\usepackage{graphicx}

\usepackage{iopams}

\begin{document}

\title[Binding energy and nature of the orbitals in fluorinated
  graphene: a DFT study]{Binding energy and nature of the orbitals in
  fluorinated graphene: a density functional theory study}

\author{F Marsusi$^{1*}$, N D Drummond$^2$}
\address{$^1$Department of Energy Engineering and Physics, Amirkabir
  University of Technology, PO Box 15875-4413, Tehran, Iran}
\ead{marsusi@aut.ac.ir}

\address{$^2$Department of Physics, Lancaster University, Lancaster
  LA1 4YB, United Kingdom} \ead{n.drummond@lancaster.ac.uk}

\vspace{10pt}
\begin{indented}
\item[]August 2017
\end{indented}

\begin{abstract}
We present density functional theory calculations of the binding
energies of one, two and three fluorine adatoms on the same side of
monolayer graphene.  We show that fluorine dimers on graphene in a
spin-singlet state are stable against dissociation into isolated
fluorine adatoms, suggesting that there is a tendency for fluorine
adatoms on a single side of graphene to cluster. Our results suggest
that fluorination develops by successive bonding of fluorine atoms to
neighbouring carbon atoms on different sublattices, while the spins
are arranged to reduce the total magnetisation of the ground state. We
find that the finite-size error in the binding energy of a single
fluorine atom or dimer on a periodic supercell of graphene scales
inversely with the cube of the linear size of the simulation
supercell. By using $\pi$-orbital axis analysis, the rehybridisation
of the three $\sigma$-orbitals pointing directly along the bonds to
the central fluorinated carbon is found to be
sp\textsuperscript{2.33}. The rehybridisation of the carbon orbital in
the C--F bond is found to be sp\textsuperscript{4.66}.
\end{abstract}

\pacs{81.05.ue, 68.35.Np, 71.15.Mb}

\vspace{2pc}
\noindent{\it Keywords}: Graphene, fluorination

\submitto{\JPCM}

\section{Introduction}
 \label{intro}

The fabrication of logic circuits based on graphene requires the
ability to open a gap in its electronic band
structure. Functionalising graphene with adatoms and admolecules is
therefore of significant interest for the development of
graphene-based electronic applications \cite{Han,Boukhvalov}. Adatoms
can significantly perturb the electronic structure of graphene,
leading to the formation of mid-gap states and the extreme
modification of the opto-electronic and transport properties. However,
to be practicable, graphene functionalised with adatoms or admolecules
of interest should be stable, even at high temperatures. Hydrogen is
one possible candidate for the band-gap engineering of graphene
\cite{Yi}. However, hydrogenated graphene suffers from instability at
moderate temperatures, restricting its applicability \cite{Yi}. On the
other hand, fluorine is a particularly attractive adatom, and it has
been confirmed that the thermal stability of fluorinated graphene is
even higher than that of pristine graphene \cite{Feng}. Previously, it
has been shown that a single fluorine adatom prefers to sit directly
above a carbon atom \cite{Nair2010,Sahin,Casolo}. According to the
literature, the binding energy $\Delta E_{\rm B}$ of a fluorine adatom
on graphene is significantly larger than that of many other adatoms
\cite{Sahin,Nakada}. While the binding energy of a single adatom is
important as a measure of the stability of fluorinated graphene, many
of its physical properties depend on the geometry and arrangement of
multiple fluorine adatoms. In this work we perform first-principles
density functional theory (DFT) calculations to investigate the
binding energy and atomic and electronic structure of a group of two
or three fluorine adatoms on graphene. This information will allow the
subsequent investigation of the thermodynamics of the fluorination
process.  We focus on the case in which the fluorine adatoms are on
the same side of the graphene layer.

We have calculated the binding energies of single adatoms and pairs of
adatoms (dimers) on the same side of $m \times m$ supercells of
monolayer graphene subject to periodic boundary conditions, where
$m=2$, 3, 4, 5, 6 and 7.  Population analysis confirms that the adatoms
set up an electric dipole moment along the C--F bond, suggesting that
finite-size effects in the binding energy due to the repulsive
interaction between the images of the dipole moments go as $-L^{-3}$,
where $L$ is the linear size of the supercell.  We have used our
results to compute the energy required to separate a fluorine dimer on
one side of graphene into two isolated single adatoms in the dilute
limit, and hence have shown that single-side dimer-fluorinated
graphene is expected to be stable.

By calculating the binding energies of single and multiple fluorine
adatoms, we show how fluorination is established geometrically on a
single side of graphene and how the electron spins are arranged in the
ground state of fluorinated graphene to increase the binding energy
and to reduce the magnetisation $M$ of the structure. Our findings are
in agreement with the experimental observation that the measured
number of paramagnetic centres is three orders of magnitude less than
the number of fluorine adatoms in fluorinated graphene samples
\cite{Nair2012}.

\section{Computational details}
\label{computational_details}

The optimised geometries and binding energies of fluorine adatoms on
monolayer graphene were obtained within the plane-wave--pseudopotential
DFT framework implemented in the \textsc{abinit} code
\cite{ABINIT1}. Both the local density approximation (LDA) and the
Perdew--Burke--Ernzerhof (PBE) \cite{Perdew} exchange--correlation
functionals were used.  The cutoff energy on the plane-wave basis set
was 40 Ha, and all atomic positions and in-plane lattice vectors were
relaxed until the atomic forces were less than 6 meV/{\AA}\@. A vacuum
region of about 19 {\AA} along the $z$ axis was imposed to guarantee a
vanishing interaction between the periodically repeated images of the
graphene layer. The binding energies of a single fluorine adatom and a
pair of fluorine adatoms on graphene were calculated in supercells of
different size, to allow extrapolation to the dilute limit. A $5\times
5\times 1$ Monkhorst--Pack \textbf{k}-point mesh was used to sample
the Brillouin zone \cite{Monkhorst} in the largest supercells, with a
finer sampling for smaller supercells. Norm-conserving
Troullier--Martins pseudopotentials were used to represent the atomic
cores \cite{Troullier,ABINIT2}. All our calculations were performed
with spin-polarised wave functions, unless otherwise stated.

\section{Results and discussion}
\label{results}

\subsection{Optimised geometries and nature of the orbitals
\label{sec:opt_geom}}

\begin{figure}
\caption{\label{FIG1} (a) Single fluorine adatom on graphene. (b) Two
  fluorine adatoms (dimer) on graphene.}
\hspace*{0.1cm}
\centering
\includegraphics[width=\textwidth]{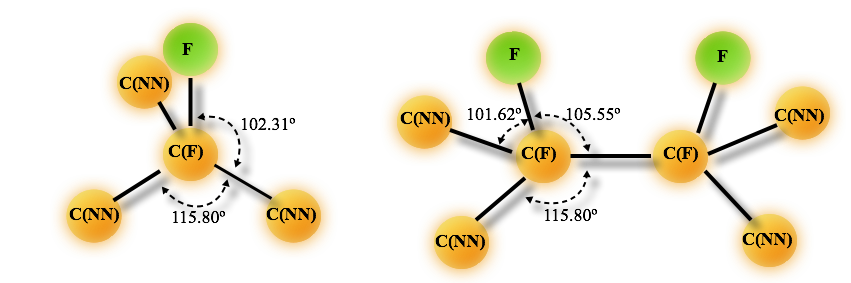}
\end{figure}

The LDA- and PBE-optimised geometrical parameters of a single fluorine
adatom and a pair of fluorine adatoms on different supercells of
graphene are shown in tables \ref{tab1}, \ref{tab2} and
\ref{tab3}. The atomic structure models are illustrated in figure
\ref{FIG1}.
\begin{table}[!pb]
\centering
\caption{\label{tab1} LDA-optimised geometry and binding energy of a
  single fluorine adatom on graphene for different supercells
  ($m\times m$).  $z$(C(F)) and $z$(C(NN)) are the $z$-coordinates of
  the carbon atom C(F) bonded to the fluorine and its nearest
  neighbour C(NN), respectively. C(F)--F and C(F)--C(NN) denote the
  bond lengths of the fluorine adatom to the carbon atom that it sits
  above and the bond length from that carbon atom to one of its
  nearest neighbours, respectively. $\angle$C(NN),C(F),F denotes the
  tetrahedral angle and $\angle$C(NN),C(F),C(NN) denotes the hexagonal
  internal angle. } \renewcommand{\arraystretch}{1.2}
\setlength\tabcolsep{5pt}
\begin{tabular}{@{}lccc@{}}
\br
& $3\times 3$ cell & $6\times 6$ cell & $7\times 7$ cell \\
\mr
$z$(C(F))${}-z$(C(NN)) & $0.32$ {\AA} & $0.32$ {\AA} & $0.32$ {\AA} \\
C--F & $1.53$ {\AA} & $1.52$ {\AA} & $1.54$ {\AA} \\
C(F)--C(NN) & $1.47$ {\AA} & $1.47$ {\AA} & $1.47$ {\AA} \\
$\angle$C(NN),C(F),F & $102.44^{\circ}$ & $102.40^{\circ}$ &
$102.26^{\circ}$ \\
$\angle$C(NN),C(F),C(NN) & $115.49^{\circ}$ & $115.52^{\circ}$ &
$115.62^{\circ}$ \\
$\Delta E_{\rm B}$ & $2.32$ eV & $2.36$ eV \\

\br
\end{tabular}
\end{table}
\begin{table}[!pb]
\centering
\caption{\label{tab2} PBE-optimised geometry and binding energy of a
  single fluorine adatom on graphene for different supercells
  ($m\times m$).  $z$(C(F)) and $z$(C(NN)) are the $z$-coordinates of
  the carbon atom C(F) bonded to the fluorine and its nearest
  neighbour C(NN), respectively. C(F)--F and C(F)--C(NN) denote the
  bond lengths of the fluorine adatom to the carbon atom that it sits
  above and the bond length from that carbon atom to one of its
  nearest neighbours, respectively. $\angle$C(NN),C(F),F denotes the
  tetrahedral angle and $\angle$C(NN),C(F),C(NN) denotes the hexagonal
  internal angle.  For comparison with a fluorine adatom, we also
  present the value of the binding energy $\Delta E_{\rm B}$ of a
  single hydrogen adatom in a $3\times 3$ cell, together with the
  corresponding value obtained in a previous work \cite{Casolo} using
  the PBE functional.}  \renewcommand{\arraystretch}{1.2}
\setlength\tabcolsep{5pt}
\begin{tabular}{@{}lcccc@{}}
\br & $3\times 3$ cell & $4\times 4$ cell & $6\times 6$ cell &
$7\times 7$ cell \\ \mr
$z$(C(F))${}-z$(C(NN)) & $0.31$ {\AA} & $0.32$ {\AA} & $0.31$ {\AA} & $0.30$ {\AA}
\\
C--F & $1.55$ {\AA} & $1.55$ {\AA} & $1.56$ {\AA} & $1.58$ {\AA} \\
C(F)--C(NN) & $1.48$ {\AA} & $1.48$ {\AA} & $1.48$ {\AA} & $1.47$
{\AA} \\
$\angle$C(NN),C(F),F & $102.46^{\circ}$ & $102.52^{\circ}$ &
$102.31^{\circ}$ & $101.96^{\circ}$ \\
$\angle$C(NN),C(F),C(NN) & $115.47^{\circ}$ &$114.44^{\circ}$ &
 $115.56^{\circ}$ & $115.82^{\circ}$ \\
$\Delta E_{\rm B}$ & $1.87$ eV & $1.93$ eV & $1.94$ eV \\
 $\Delta E_{\rm B}$ (Hydrogen) & $0.76$, $0.77$ eV\,$^a$&\multicolumn{3}{c}{} \\
\br
\end{tabular}\\
$^{a}$From reference \cite{Casolo}.
\end{table}
\begin{table}[!pb]
\centering
\caption{\label{tab3} LDA- and PBE-optimised geometries of two
  same-side fluorine adatoms (singlet state) on graphene for different
  supercell sizes: $3\times 3$ and $6\times 6$. The graphene monolayer
  before adsorption lies in the plane $z=0$; the centre of mass of the
  unit cell is pinned during relaxation. $z$(C(F)) and $z$(C(NN)) are
  the $z$-coordinates of each carbon C(F) bonded to fluorine and its
  nearest neighbour C(NN), respectively. The bond lengths of C(F)--F
  and C(F)--C(NN) are shown.  Values of $\Delta U_{\rm B}$ and second
  adsorption energy ($\Delta E_{\rm B2}=E({\rm G}+{\rm F})+E({\rm
    F})-E({\rm G}+2{\rm F})$) of two hydrogen adatoms on a $3\times 3$
  supercell of graphene using the PBE functional obtained in this
  work, as well as $\Delta E_{\rm B2}$ obtained in a previous work
  \cite{Casolo} using the PBE functional in a $5\times 5$ cell are
  shown for comparison with the results of two fluorine adatom on
  graphene.  \@.}  \renewcommand{\arraystretch}{1.2}
\setlength\tabcolsep{5pt}
\begin{tabular}{@{}lcccc@{}}
\br

& \multicolumn{2}{c}{LDA} & \multicolumn{2}{c}{PBE} \\

& $3\times 3$ cell & $6\times 6$ cell & $3\times 3$ cell & $6\times 6$
cell \\
\mr

$z$(C)${}-z$(C(NN)) & $0.48$ {\AA} & $0.50$ {\AA} & $0.49$ {\AA} & $0.50$ {\AA} \\

C--F & $1.44$ {\AA} & $1.45$ {\AA} & $1.53$ {\AA} & $1.45$ {\AA} \\

C(F)--C(F) & $1.54$ {\AA} & $1.56$ {\AA} & $1.56$ {\AA} & $1.57$ {\AA} \\

C(NN)--C(F) & $1.49$ {\AA} & $1.49$ {\AA} & $1.50$ {\AA} & $1.50$ {\AA} \\

$\angle$C(NN),C(F),C(F) & $115.81^{\circ}$ & $115.80$ &
$115.62^{\circ}$ & $115.64^{\circ}$ \\

$\angle$F,C(F),C(F) & $105.59^{\circ}$ & $105.55^{\circ}$&
$105.89^{\circ}$ & $105.84^{\circ}$ \\

$\angle$F,C(F),C(NN) & $101.27^{\circ}$ & $101.62^{\circ}$ &
$101.58^{\circ}$ & $101.80^{\circ}$ \\

$\Delta U_{\rm B}$ & $0.72$ eV & $0.71$ eV & $0.68$ eV & $0.65$ eV \\

$\Delta U_{\rm B}$ (Hydrogen) & & & $1.16$ eV & \\

$\Delta E_{\rm B2}$ (Hydrogen) & & & \multicolumn{2}{c}{$1.92$, $1.93$
  eV$\,^a$} \\
\br
\end{tabular} \\
$^{a}$From reference \cite{Casolo}.
\end{table}

After full relaxation, a single fluorine adatom remains exactly on top
of a carbon atom, as expected. However, the repulsive force between
two fluorine adatoms on top of neighbouring carbon atoms makes them
relocate from their initial positions, as shown in figure
\ref{FIG1}(b). The geometry around the fluorine adatom appears to be
converged with respect to size in a $7\times 7$ supercell. Let C(F)
denote the carbon atom to which the fluorine adatom is bonded. In a
single adatom on graphene, the C(F)--F bond length is predicted to be
about 1.55 {\AA}, which is larger than the typical C--F sp$^3$ bond
length (about 1.37 {\AA}), and is in good agreement with a previous
PBE prediction \cite{Santos}. Other geometric information, including
the distance between C(F) and its nearest neighbours C(NN), is
presented in tables \ref{tab1}, \ref{tab2} and \ref{tab3}. As a
consequence of the attractive interaction between the carbon and
fluorine atoms, C(F) is pulled out of the plane by about 0.33 {\AA}
and shows a strong tendency to form a near sp$^3$ hybridisation. The
three C(F)--C(NN) $\sigma$-bonds of graphene resist stretching;
consequently, these atoms are also slightly dragged out of the plane
to reduce the stress over bonds. This results in a reduction of the
internal angles $\angle$C(NN),C(F),C(NN) in the hexagonal ring around
C(F) from $120.0^{\circ}$ to about $115.5^{\circ}$.

We have performed $\pi$-orbital axis vector (POAV1) analysis as
described in references \cite{Haddon_Chem} and \cite{Bai} to evaluate
the nature of the orbitals in fluorinated graphene. This method is
based on the coordinates of the conjugated central carbon atom C(F)
and the three neighbouring carbon atoms C(NN)\@. POAV1 predicts a
deviation from sp$^2$ to sp$^{2.33}$ hybridisations for the three
C(F)--C(NN) $\sigma$ bonds for a single fluorine adatom. POAV2
analysis can be used to predict the nature of the C--F bond by
calculating the degree of the p content in the $\sigma$ orbitals
(sp$^n$) \cite{Haddon_Pure}. The three $\sigma$-bonds resist further
pulling up of the C(F)\@. This results in a larger C--F bond length
and causes $\angle$F,C(F),C(NN) to be $102.4^{\circ}$ compared with
the $109.5^{\circ}$ characteristic of sp$^3$ bonding. Analysing the
$\sigma$- and $\pi$-orbitals of atoms using the POAV2 method
demonstrates the formation of sp$^{4.66}$ rehybridisation in the
C(F)--F bond, which indicates the great contribution of nearby
p$_z$-orbitals to this bond. Forming sp$^{4.66}$ hybridisation, rather
than sp$^3$, arises due to the large electro-negativity of fluorine
together with the tendency of the $\sigma$-bonds to maintain sp$^2$
hybridisation in the graphene sheet.

In our study of the most stable geometry for two fluorine adatoms on
the same side of a graphene sheet, many possible configurations have
been investigated. From our calculations of the binding energy $\Delta
E_{\rm B}$, which will be presented in section
\ref{sec:binding_energies}, we have found that the most stable
geometry is the structure with two fluorine adatoms above two
neighbouring carbon atoms. This is about 0.87 eV more stable than the
structure with two fluorine adatoms above next-nearest neighbouring
carbon atoms. If we denote the two sublattices of graphene's bipartite
hexagonal lattice as A and B, this would mean the most stable bonding
occurs when the two adatoms are bonded to two sites of opposite
sublattice A and B\@. A similar conclusion was reached in a PBE study
of the adsorption of a second hydrogen adatom on graphene
\cite{Casolo}.

A fluorine atom has an unpaired electron, which carries a spin moment
of 1 $\mu_{\rm B}$ magnetisation, where $\mu_{\rm B}$ is the Bohr
magneton. For a fluorine dimer on graphene, two different states with
singlet and triplet spin arrangements are possible. In the triplet
case, the two parallel spins further avoid each other in accordance
with the Pauli exclusion principle, leading to a larger distance
between the two fluorine adatoms and causing a larger stretching of
the C(F)--F bond (1.47 against 1.44 {\AA}), and consequently
decreasing the stability of the structure by 1.5 eV\@. The C(NN)--C(F)
bond length of the singlet state was calculated to be 1.49 {\AA},
which is larger than the corresponding value of 1.47 {\AA} for a
single fluorine adatom. C.f., the C--C bond length in graphene is
about 1.42 {\AA}\@. In the case of two fluorine adatoms, each fluorine
atom repels the other, pulling the C(F) atoms further out of the plane
by about 0.45 {\AA}\@. This is a sign of more stress on the three
C--C(F) $\sigma$ bonds compared with the single-adatom case.

\subsection{Binding energies: single and two adatoms
\label{sec:binding_energies}}

\begin{figure}
\caption{\label{FIG2} PBE binding energy of a single fluorine adatom
  on an $m\times m$ ($m=2$, 3, 4 and 6) periodic supercell of graphene
  (black squares). The binding energy is extrapolated to the dilute
  limit of infinite cell size ($L\longrightarrow\infty$) by fitting
  equation (\ref{eq:binding_fit}) to the data. The fitted parameter
  values are $\Delta E_{\rm B}(\infty)=1.95$ eV and $c=-1.82$
  eV\,$a_0^{3}$, with the root-mean-square error being 0.02 eV\@.} \centering
\includegraphics[width=0.8\textwidth]{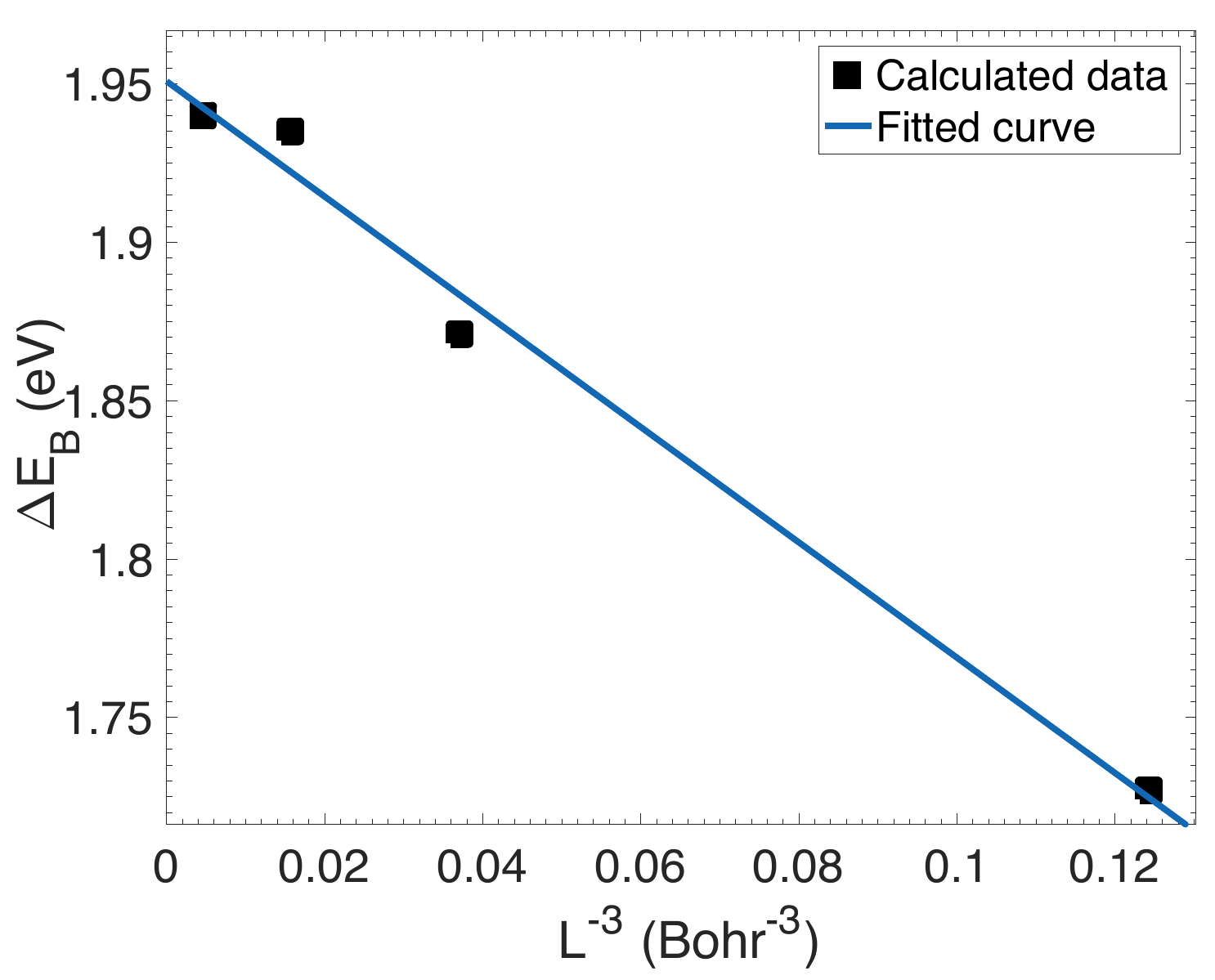}
\end{figure}

The binding energy $\Delta E_{\rm B}$ and formation energy $\Delta
E_{\rm F}$ of a single adatom or of multiple adatoms on a graphene
sheet are difficult to measure directly due to the small size of the
defects. Instead, DFT is an ideal tool to provide quantitative
estimates of $\Delta E_{\rm B}$ and $\Delta E_{\rm F}$ at low
concentrations. $\Delta E_{\rm B}$ is defined as the energy required
to separate a single, isolated adatom from the graphene surface to
infinity:
\begin{equation}
\Delta E_{\rm B}=E({\rm G})+E({\rm F})-E({\rm G}+{\rm F}), \label{eq:E_B}
\end{equation}
where $E({\rm F})$ is the total energy of an isolated fluorine
atom. $E({\rm G})$ and $E({\rm G}+{\rm F})$ are the total energies of
a large graphene sheet and a singly fluorinated graphene sheet,
respectively.  $\Delta E_{\rm B}$ is computed to investigate the
stability and the strength of the C--F covalent bond of a single
fluorine adatom on graphene. We also introduce $\Delta U_{\rm B}$ as
the energy required to separate two fluorine adatoms to infinite
distance from each other on a graphene sheet. The formation energy $\Delta
E_{\rm F}$ per adatom of fluorinated graphene relative to pristine
graphene and the F$_2$ free molecule is another quantity that
characterises the stability of single-side fluorinated graphene
\cite{Yi,Sofo,Leenaerts}. From these definitions, $\Delta U_{\rm B}$
and $\Delta E_{\rm F}$ can be calculated by:
\begin{eqnarray}
\Delta U_{\rm B} & = & E({\rm G})+2E({\rm F})-2\Delta E_{\rm B}-E({\rm
  G}+2{\rm F}) \nonumber \\ & = & 2E({\rm G}+{\rm F})-E({\rm
  G})-E({\rm G}+2{\rm F}). \label{eq:U_B} \end{eqnarray} and
\begin{equation}
\Delta E_{\rm F} = [E({\rm G})+E({\rm F}_2)-E({\rm G}+2{\rm
    F})]/2, \label{eq:E_F}
\end{equation}
where $\Delta E_{\rm B}$ is given by equation (1). $E({\rm G}+2{\rm
  F})$ is the total energies of a graphene sheet with a pair of
fluorine adatoms placed in stable positions roughly above two
neighbouring carbon sites (see figure \ref{FIG2}(b)) and $E({\rm
  F}_2)$ is the total energy of a free F$_2$ molecule. Special care
was taken to ensure that the same ${\bf k}$ points were used in the
graphene and fluorinated graphene calculations. For example, one must
use a $15\times 15 \times 1$ ${\bf k}$-point grid centred on $\Gamma$
for a primitive cell of graphene to compare its energy $E({\rm G})$
(scaled up to the supercell size) with the energy of fluorinated
graphene $E({\rm G}+{\rm F})$ in a $3\times 3$ supercell with a
$5\times 5 \times 1$ ${\bf k}$-point grid centred on $\Gamma$.

From the definitions given in equations
(\ref{eq:E_B})--(\ref{eq:E_F}), a \textit{negative} value for $\Delta
E_{\rm B}$, $\Delta U_{\rm B}$ or $\Delta E_{\rm F}$ would indicate
that the fluorine dimer on graphene is unbound. The predicted values
of $\Delta E_{\rm B}$ and $\Delta U_{\rm B}$ at various cell sizes are
shown in tables \ref{tab1}, \ref{tab2} and \ref{tab3}. The PBE values
of $\Delta E_{\rm B}$, $\Delta E_{\rm F}$ and $\Delta U_{\rm B}$
obtained in a $6\times 6$ supercell are 1.95, 1.07 and 0.65 eV,
respectively, while the corresponding LDA values are 2.36, 1.20 and
0.71 eV\@.

The binding energies $\Delta E_{\rm B}$ of single fluorine adatoms on
$3\times 3$ and $4\times 4$ supercells of graphene were calculated in
reference \cite{Liu} using the Perdew--Wang (PW) generalised gradient
approximation (GGA) exchange--correlation functional \cite{Perdew1992}
using the projector-augmented-wave (PAW) method, but the value
obtained in the dilute limit is about 0.41 eV higher than our PBE
results. We believe the difference between our results and those of
reference \cite{Liu} is due to the small plane-wave cutoff (18.37
a.u.)\ applied in that work. We found that a plane-wave cutoff energy
of 28 a.u.\ is required for adequate convergence within the PAW
method. We verified that the GGA-PBE functional over-binds by a similar
amount (binding energy of 2.23 eV) when a cutoff energy of 18.37
a.u.\ is used.

Finite-size effects are a systematic source of errors in our
calculations. Finite-size errors arise from the non-physical and
unwanted interactions between the periodic images of the adatoms
within the plane of the graphene sheet. L{\"o}wdin population analysis
\cite{Lowdin} shows that a relatively large charge of about $0.38|e|$
is transferred from C(F) to the fluorine adatom, giving ionic
character to the C(F)--F covalent bond and causing the defect to have
a nonzero electric dipole moment.  The unwanted electrostatic energy
of a 2D lattice of identical dipole moments is positive and falls off
as $L^{-3}$, where $L$ is the linear size of the cell \cite{Makov}.
The function
\begin{equation}
\Delta E_{\rm B}(L)=\Delta E_{\rm B}(\infty)+cL^{-3}, \label{eq:binding_fit}
\end{equation}
was fitted to our PBE results, where $\Delta E_{\rm B}(L)$ is the
$\Delta E_{\rm B}$ value in a cell of linear size $L$. The
extrapolation is shown in figure \ref{FIG2}.  Note that the positive
error in $E({\rm G}+{\rm F})$ results in $c$ being negative.  The
binding energy $\Delta E_{\rm B}(\infty)$ in the dilute limit is
$1.95$ eV\@.

As can be seen in figure \ref{FIG2}, there are also non-systematic
finite-size effects in the binding energy due to the choice of ${\bf
  k}$-point sampling, long-range oscillations in the charge density
around the defect and the strain field resulting from the geometric
distortion of the graphene lattice due to the fluorine defect.  In
general, as the supercell size is increased, the Brillouin zone to be
sampled shrinks and hence the required number of ${\bf k}$ points
decreases in inverse proportion to the supercell size.  The results in
tables \ref{tab4} and \ref{tab5} indicate that the binding energy
$\Delta E_{\rm B}$ converges to better than 10 meV with a $7\times
7\times 1$ Monkhorst--Pack \textbf{k}-point mesh in a $3\times 3$
fluorinated graphene supercell, while for a larger $4\times 4$
supercell it is converged to this level with a $5\times 5\times 1$
\textbf{k}-point grid. The bond lengths and angles at each supercell
size are listed in tables \ref{tab1}, \ref{tab2} and \ref{tab3}. These
show that the atomic structure in the vicinity of the defect has
converged to a high degree of precision in a $6\times 6$ supercell,
suggesting that finite-size errors in the binding energy due to
geometrical effects may be small.

\begin{table}[!pb]
\centering
\caption{\label{tab4} Relative total energies $\Delta E({\rm G}+{\rm
    F})$ and binding energies of a single fluorine adatom on a $m
  \times m$ supercell of graphene. Data are obtained using different
  Monkhorst--Pack ${\bf k}$-point grids in the Brillouin zone and
  supercells with different size.  Here, $\Delta E({\rm G}+{\rm F})=
  E({\rm G}+{\rm F})_{k\times k\times1}-E({\rm G}+{\rm F})_{l\times
    l\times1}$, in which the numbers $k$ and $l$ represent the ${\bf
    k}$-point grid used in the row numbers $n+1$ and $n$ ($n=1\ldots
  5$) of this table, respectively. } \renewcommand{\arraystretch}{1.2}
\setlength\tabcolsep{5pt}
\begin{tabular}{@{}lccccc@{}}
\br
 & \multicolumn{2}{c}{$\Delta E({\rm G}+{\rm F})$ (meV)} &
\multicolumn{2}{c}{$\Delta E_{\rm B}$ (eV)} \\
\raisebox{1.5ex}[0pt]{${\bf k}$-point mesh} & $3\times 3$ cell & $4\times
4$ cell & $3\times 3$ cell & $4\times 4$ cell \\
\mr
$3\times 3\times 1$ & $0.00$ & $0.00$ & $2.04$ & $2.04$ \\
$5\times 5\times 1$ & $21.28$ & $-2.72$ & $1.94$ & $1.93$ \\
$7\times 7\times 1$ & $-17.42$ & $-5.44$ & $1.90$ & $1.94$ \\
$9\times 9\times 1$ & $3.86$ & $5.44$ & $1.90$ & $1.95$ \\
$11\times 11\times 1$ & $3.62$ & $0.00$ & $1.90$ & $1.94$ \\

\br
\end{tabular}
\end{table}
\begin{table}[!pb]
\centering
\caption{\label{tab5} Relative total energies $\Delta E({\rm G}+{\rm
    F})$ and binding energies of a single fluorine adatom on a
  supercell of graphene. Data are obtained from different
  Monkhorst--Pack grids in the Brillouin zone of a $6\times 6$
  supercell. Here, $\Delta E({\rm G}+{\rm F})= E({\rm G}+{\rm
    F})_{k\times k\times1}-E({\rm G}+{\rm F})_{l\times l\times1}$, in
  which the numbers $k$ and $l$ represent the ${\bf k}$-point grid
  used in the row numbers $n+1$ and $n$ ($n=1,2,3$) of this table,
  respectively.}  \renewcommand{\arraystretch}{1.2}
\setlength\tabcolsep{5pt}
\begin{tabular}{@{}lcc@{}}
\br
${\bf k}$-point mesh & $\Delta E({\rm G}+{\rm F})$ (meV) & $\Delta
E_{\rm B}$ (eV) \\
\mr
$3\times 3\times 1$ & $0.00$ & $2.03$ \\
$5\times 5\times 1$ & $8.16$ & $1.97$ \\
$7\times 7\times 1$ & $2.72$ & $1.96$ \\
\br
\end{tabular}
\end{table}
As described in section \ref{sec:opt_geom}, and in agreement with a
previous study \cite{Casolo}, we found that two fluorine adatoms on
top of two nearest-neighbour carbon atoms belonging to the hexagonal A
and B sublattices, respectively, is the most stable single-side
two-fluorine-adatom arrangement. The spin part of the ground state can
be either a spin-paired singlet ($\uparrow\downarrow$) or a
spin-unpaired triplet ($\uparrow\uparrow$) state. In the spin-unpaired
arrangement, the two parallel spins further avoid each other, which
leads to an increase in the total energy such that both LDA and PBE
calculations predict $\Delta U_{\rm B}$ to take a negative value, as
can be seen in table \ref{tab2}. Consequently, two parallel-spin
fluorine adatoms are expected to repel each other, and clustering can
only be made from anti-parallel spins from adatoms on nearest neighbours.  
In accordance with the LDA's general tendency to over-bind compared to
the PBE functional \cite{Gross}, the absolute value of $\Delta E_{\rm
  B}$ predicted by the LDA differs by about 0.41 eV from the value
predicted by the PBE functional. The over-binding of singly
fluorinated graphene predicted by the LDA is associated with the
smaller C(F)--F bond length predicted by this functional. PBE and LDA
converge to the same C(F)--F bond length of 1.45 {\AA} for a fluorine
dimer on graphene; correspondingly the predicted $\Delta U_{\rm B}$
from LDA and PBE calculations are almost the same (the LDA $\Delta
  U_{\rm B}$ value is only 0.02 eV lower than the PBE value).

The electronic behaviour of fluorine adatoms on graphene differs from
that of hydrogen adatoms.  For the second \textit{hydrogen} added to
graphene, we find that $\Delta U_{\rm B}=1.16$ eV, which is even
larger than the first adsorption energy ($\Delta E_{\rm B}=0.76$ eV),
as seen in table \ref{tab2}, and from the first and second adsorption
energies presented in a previous study \cite{Casolo}. In fact, the
second adatom couples to the unpaired electron available on the
nearest-neighbour site. Fluorine is a strongly electro-negative
element and has a great tendency to bind to carbon. Unlike hydrogen,
however, there is a significant repulsive interaction between pairs of
fluorine adatoms on neighbouring sites due to the overlap between
their relatively delocalised orbitals, which decreases $\Delta U_{\rm
  B}$. Nevertheless, the experimental F--F bond energy (1.61 eV) is
significantly smaller than the C--F bond energy (5.03 eV); in
contrast, the H--H bond energy (4.48 eV) is larger than that of C--H
(4.26 eV) \cite{National}. Hence fluorinated graphene is more stable
than hydrogenated graphene, despite the repulsive interaction between
the fluorine atoms on neighbouring carbon atoms.

The LDA and PBE formation energies of a single fluorine adatom on a
$3\times 3$ supercell of graphene are about $\Delta E_{\rm F}=1.17$
and 0.99 eV, respectively. These values hardly change as the cell size
is increased. The \textit{positive} formation energy and $\Delta
U_{\rm B}$ indicate that single-side fluorinated graphene is stable,
while these values are about two orders of magnitude higher than
typical room temperature ($\sim26$ meV)\@. This thermal stability
makes fluorinated graphene distinctive from hydrogenated graphene and
potentially more suitable for electronic and spintronic
applications. Interestingly, while the binding energies of
hydrogenated and fluorinated graphene are not so different, there is a
significant difference in their formation energy. The formation energy
is a measure of stability against molecular desorption from the
graphene surface. In contrast to fluorinated graphene, hydrogenated
graphene readily dissociates into graphene and H$_2$ molecules
\cite{Yi}. The different behaviour is the result of the large
difference in the F--F and H--H bond energies, as explained in section
\ref{sec:binding_energies}.

\subsection{Binding energies: three adatoms}\label{sec:three_adatom}

\begin{figure*}
\caption{\label{FIG3} Different arrangements of three fluorine adatoms
  on graphene. The bonding site is labelled as A or B, depending on
  the associated sublattice. The prime denotes an arrangement with a
  third adatom on a non-neighbouring carbon atom.}
\hspace*{0.4 cm}
\centering
\includegraphics[width=\textwidth]{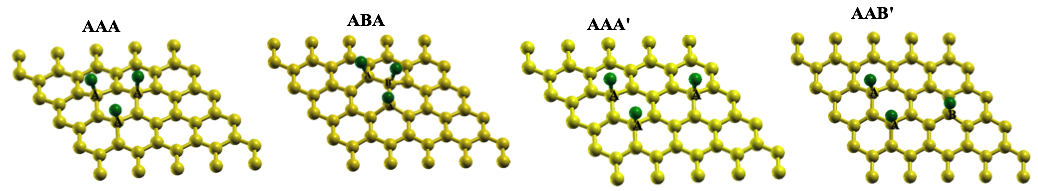}
\end{figure*}

In section \ref{sec:binding_energies} we found that, for a second
fluorine adatom on graphene, it is energetically favourable to keep
the two sublattices in balance, with the least possible
magnetisation. To understand further the fluorination process, we have
studied the behaviour of three fluorine adatoms on $3\times 3$ and
$5\times 5$ supercells of graphene. We have applied the general
definition of the binding energy per adatom:
\begin{equation}
\overline{\Delta E_{\rm B}}=\frac{E({\rm G})+n_{\rm F}E({\rm F})-E({\rm G}+n_{\rm
    F}{\rm F})}{n_{\rm F}},
\end{equation}
where $n_{\rm F}$ is the number of fluorine adatoms. We have investigated
the different possible configurations of the three adatoms on top of
the carbon atoms in a hexagonal ring, as shown in figure
\ref{FIG3}. We refer to the different arrangements of fluorine atoms
shown in figure \ref{FIG3} as AAA, ABA, AAA$'$ and AAB$'$.  We find
that binding the three fluorine atoms to three carbons from the same
sublattice in the AAA arrangement is the least favorable configuration
($\overline{\Delta E_{\rm B}}=1.69$ and 1.60 eV per fluorine adatom in
a $3\times 3$ and $5\times 5$ cell, respectively). With the exception
of the AAA arrangement, the $z$-coordinates of the three C(F) atoms
differ by about 0.1 {\AA}. In the ABA structure, the $z$-coordinate of
C(F) is larger at the B site; in the AAB$'$ and AAA$'$ structures the
$z$-coordinate of the isolated C(F) atom is smaller than that of the
other two C(F) atoms.

Binding-energy results are listed in tables \ref{tab6} and
\ref{tab7}. As in the adsorption of two adatoms, it is energetically
favourable to reduce the imbalance between the two sublattices by
adsorption on mixed sites. Correspondingly, ABA is the most favoured
arrangement by about 1 eV in both supercells. The C--F bond length
also is the least for the ABA arrangement. After adsorption of two
adatoms on different sites of the A and B sublattices, there is no
significant preference in the bonding of a third adatom to a carbon
placed in the A or B sublattice. For example, the difference between
$\overline{\Delta E_{\rm B}}$ in ABA ($\overline{\Delta E_{\rm
    B}}=2.10$ eV per fluorine adatom) and ABB ($\overline{\Delta
  E_{\rm B}}=2.09$ eV per fluorine adatom) arrangements in the $3\times$3
supercell is small. The larger $5\times$5 cell allows us to inspect the
change in binding energy when the third adatom is far from the first
ad-dimer. We found that the third fluorine in a mixed-site
arrangement, such as AAB$'$, is favored over the same sites
arrangements AAA or AAA$'$; see tables \ref{tab7} and
\ref{FIG3}. Moreover, we compared the energy required to separate
three adatoms from AAB and AAA configuration to infinite distance from
each other using the general equation:
\begin{equation}
\overline{\Delta U_{\rm B}} = E({\rm G})+n_{\rm F} E({\rm F})-n_{\rm
  F}\Delta E_{\rm B}-E({\rm G}+n_{\rm F}{\rm F}),
   \label{eq:U_B_gen} 
\end{equation}
where $n_{\rm F}$ is the number of fluorine adatoms. The results are
listed in table \ref{tab7}. We observe that, in contrast to ABA, the
AAA arrangement has a negative $\overline{\Delta U_{\rm B}}$ value and
is therefore not a stable arrangement. These outcomes, along with the
previous results obtained for an ad-dimer in the section
\ref{sec:binding_energies}, suggest that fluorination develops
geometrically from a central carbon, with additional fluorine adatoms
bonding to neighbouring carbon atoms, and so on.

 Another interesting result is that the value of the total
 magnetisation decreases with increasing binding energy
 $\overline{\Delta E_{\rm B}}$, as predicted by the PBE functional. In
 general, semi-local functionals such as PBE cannot reliably predict
 the total magnetisation of a fluorine adatom on graphene \cite{Kim},
 and more sophisticated methods are needed for quantitative
 results. The non-integer total magnetisations presented in tables
 \ref{tab6} and \ref{tab7} are due to the need for a more advanced
 level of theory than DFT-GGA\@. However, at least we can use the
 prediction of the PBE functional to compare qualitatively the trend
 of the change in the total magnetisation in terms of the binding
 energies. According to PBE, the total magnetisation $M$ is the least for
 the most stable arrangements listed in tables \ref{tab6} and
 \ref{tab7}.
\begin{table}[!pb]
\centering
\caption{\label{tab6} Relative total energy of adatoms on graphene in
  a 3$\times$3 supercell (compared to the total energy of the AAA
  configuration of adatoms), binding energies per adatom, PBE magnetic
  moment in units of $\mu_{\rm B}$, $|n_{\rm A}-n_{\rm B}|/2$, and the
  energy of the unpaired-spin configuration relative to the
  paired-spin configuration: $\Delta E_{\rm
    S}=E(\uparrow\uparrow\uparrow)-E(\uparrow\downarrow\uparrow)$. $n_{\rm
    A}$ and $n_{\rm B}$ are the numbers of fluorine adatoms above
  carbon atoms on the A and B hexagonal sublattices, respectively.}
\renewcommand{\arraystretch}{1.2} \setlength\tabcolsep{5pt}
\begin{tabular}{@{}lccccc@{}}
\br
Configuration & $\Delta E({\rm G}+3{\rm F})$ (eV) & $\overline{\Delta E_{\rm B}}$ (eV)
& $M$ ($\mu_{\rm B}$) & $|n_{\rm A}-n_{\rm B}|/2$ & $\Delta E_{\rm S}$ (eV) \\
\mr
ABA & $-1.25$  & $2.10$ & $0.28$& $1/2$ & $0.75$ \\
BAB & $-1.22$  & $2.09$ & $0.43$& $1/2$ &  \\
AAA & $0.00$  & $1.69$ & $1$& $3/2$ & \\
\br
\end{tabular}
\end{table}
\begin{table}[!pb]
\centering
\caption{\label{tab7} Relative total energy of three fluorine adatoms
  on graphene in a 5$\times$5 supercell (compared to the total energy
  of the AAA configuration of adatoms), binding energy per adatom,
  energy required to separate three fluorine adatoms to infinite
  distance, PBE predicted magnetic moment in units of $\mu_{\rm B}$ and
  $|n_{\rm A}-n_{\rm B}|/2$. The C(F)--F bond length of each
  arrangement shown in figure \ref{FIG3} is also listed.}
\renewcommand{\arraystretch}{1.2} \setlength\tabcolsep{5pt}
\begin{tabular}{@{}lccccccc@{}}
\br
Config. & $\Delta E({\rm G}+3{\rm F})$ (eV) & $\overline{\Delta E_{\rm B}}$
(eV) & $\overline{\Delta U_{\rm B}}$ (eV) & $M$ ($\mu_{\rm B}$) &$|n_{\rm A}-n_{\rm B}|/2$& C--F ({\AA}) \\
\mr
AAA      & $0.00$       & $1.60$      & $-0.34$     & $1$   &  $3/2$   &  1.52      \\
ABA      & $-0.98$      & $1.94$       & $0.64$     & $0$   &   $1/2$  &    $1.46 $    \\
AAA$^\prime$ & $0.02$  & $1.60$ & $-0.36$ & $0.5$ & $3/2$ & $1.55$ \\
AAB$^\prime$ & $-0.35$ & $1.73$ & $0.01$  & $0$   & $1/2$ & $1.53$  \\
\br
\end{tabular}
\end{table}

The total energies of three fluorine adatoms on graphene obtained in
spin-polarised calculations are typically lower than the energies
obtained in spin-unpolarised calculations by about 184 meV\@ in a
3$\times$3 cell. To get more insight into the effects of spin
configuration on the binding energy, we have also performed
magnetisation-constrained calculations for ABA, which is the most stable
configuration in a 3$\times$3 cell. Our results again show that the
configuration with two paired spins ($\uparrow\downarrow\uparrow$) is
preferred over three parallel spin ($\uparrow\uparrow\uparrow$) by
about 748 meV per adatom. This is consistent with Lieb's theorem
\cite{Lieb}, and explains the experimental observation in which the
measured number of paramagnetic centres is three orders of magnitude
less than the number of fluorine adatoms in fluorinated graphene
samples \cite{Nair2012}. Also, according to Lieb's theorem, the ground
state of a bipartite lattice with a half-filled band has a total spin
of $|n_{\rm A}-n_{\rm B}|/2$. For pristine graphene the numbers of
sites in the A and B sublattices are equal ($n_{\rm A}=n_{\rm B}$),
leaving the ground state of pristine graphene with zero net spin. For
the ABA arrangement of fluorine adatoms on graphene, for which
$n_{\rm A}=N/2-2$ and $n_{\rm B}=N/2-1$, $S$ should be equal to $1/2$,
which leaves an unpaired spin (for example $\uparrow\downarrow\uparrow$
configuration against $\uparrow\uparrow\uparrow$ spin-arrangement)
with total magnetisation $M=1~\mu_{\rm B}$. The same outcomes are
deduced when two and three hydrogen atoms are adsorbed on the graphene
sheet \cite{Casolo}. All these results lead us to conclude that when
two or more fluorine atoms with unpaired spins are adsorbed on
graphene, long-ranged spin polarisations, of opposite signs are
induced in the two sublattices to reduce the total magnetisation. So
we see that ABA and ABA$'$ with the least magnetisations are
energetically similar (ABA is preferred), and are the most stable
configurations. This interpretation from the energies point of view
has been deduced explicitly by higher-level calculations using a
hybrid functional or a DFT+U model \cite{Kim, Marsusi}.
 
\section{Conclusions}

In summary, we have studied the orbital nature of the C--F bond in
fluorinated graphene. Using the POAV method we find sp$^{4.6}$
rehybridisation for the C--F bond of a single fluorine adatom on a
graphene sheet, implying large contributions from nearby p
orbitals. The behaviour of the binding energy of a fluorine adatom on
graphene in supercells of various sizes was investigated, and the
results were compared with hydrogenated graphene within the DFT
framework. Both the LDA and PBE functionals predict that a fluorine
dimer adsorbed onto neighbouring carbon atoms is a stable structure,
suggesting that there should be a tendency for fluorine atoms to
cluster during a fluorination process in which a single side of
graphene is exposed to fluorine. Our calculations show that, in
contrast to hydrogenated graphene, the formation energy of single-side
fluorinated graphene is positive, which implies fluorinated graphene
should be more stable than hydrogenated graphene at higher
temperature. Also, for multiple fluorine adatoms the configurations in
which the fluorine atoms are bonded to sites corresponding to the same
sublattice are less stable. This suggests that fluorination proceeds
with fluorine atoms successively bonding to neighbouring carbon
atoms. The spins density of states in the ground state are arranged
according to minimise the total magnetisation.  We have shown that the
finite-size error in the binding energy of isolated fluorine adatoms
or ad-dimers stems from the interaction of the electric dipole moments
of images of the defects in neighbouring cells, and falls off as the
inverse cube of the linear cell size.

\section{Acknowledgements}

F Marsusi appreciates the computer assistance provided by Mrs Z
Zeinali in the Department of Energy Engineering and Physics at
Amirkabir University of Technology.

\end{document}